\documentclass[conference]{IEEEtran}
\usepackage{amsmath}
\usepackage{amsthm}
\usepackage{amssymb}
\usepackage{amsfonts}
\usepackage{mathrsfs}
\usepackage{algorithmic,algorithm}
\usepackage{graphicx}
\usepackage{subfigure}
\usepackage{cite}
\usepackage{bm,epsfig,amsthm,url}
\usepackage{indentfirst}
\usepackage{epstopdf}
\newtheorem{remark}{Remark}

\DeclareMathOperator*{\argmin}{arg\,min}

\begin{document}

\title{Optimal Receive Beamforming for Over-the-Air Computation}
\author{
  \IEEEauthorblockN{$\text{Wenzhi Fang}$, Yinan Zou, $\text{Hongbin Zhu}$,$\text{Yuanming Shi}$, and $\text{Yong Zhou}$
  }
 \IEEEauthorblockA{
  School of Information Science and Technology, ShanghaiTech University, Shanghai 201210, China \\ 
  Email: \{fangwzh1, zouyn, zhuhb1, shiym, zhouyong\}@shanghaitech.edu.cn}}

%


\maketitle

\begin{abstract}

In this paper, we consider fast wireless data aggregation via over-the-air computation (AirComp) in Internet of Things (IoT) networks, where an access point (AP) with multiple antennas aim to recover the arithmetic mean of sensory data from multiple IoT devices. 
To minimize the estimation distortion, we formulate a mean-squared-error (MSE) minimization problem that involves the joint optimization of the transmit scalars at the IoT devices as well as the denoising factor and the receive beamforming vector at the AP. 
To this end, we derive the transmit scalars and the denoising factor in closed-form, resulting in a non-convex quadratic constrained quadratic programming (QCQP) problem concerning the receive beamforming vector.
Different from the existing studies that only obtain sub-optimal beamformers, we propose a branch and bound (BnB) algorithm to design the globally optimal receive beamformer.
Extensive simulations demonstrate the superior performance of the proposed algorithm in terms of MSE. 
Moreover, the proposed BnB algorithm can serve as a benchmark to evaluate the performance of the existing sub-optimal algorithms.


\end{abstract}


\section{Introduction}

With the rapid advancement of smart city, internet of vehicles, and edge artificial intelligence,
it is expected that Internet of Things (IoT) will support ubiquitous connectivity for billions of devices that generate massive amount of real-world data \cite{zhu2020overtheair}.
Wireless data aggregation among the distributed IoT devices is an important but challenging task \cite{blind,wang_iot,wang2020federated}. 
Due to the scarcity of spectrum resources and the ultra-low latency requirement,  the conventional transmit-then-compute scheme cannot support fast wireless data aggregation in dense IoT networks. 
Fortunately, over-the-air-computation (AirComp) has the potential to achieve fast wireless data aggregation by enabling the paradigm of ``compute when communicate". 
In particular, by exploiting the superposition property of multiple access channels (MACs), wireless data aggregation can be achieved in one transmission interval by allowing all IoT devices to transmit concurrently over the same radio channel \cite{yangkai}. 

%


AirComp was firstly investigated in the seminal work \cite{info}, where the authors showed that the superposition property of MACs can be exploited to compute the nomographic functions from an information theoretical perspective.
With the great potential for wireless data aggregation, AirComp has recently attracted considerable interests \cite{optimal_liu, optimal_huang, chen2018uniform, multi_function, Muti_modal}.
In particular, considering simple single-input single-output (SISO) wireless networks with energy-constrained IoT devices, the authors in \cite{optimal_liu, optimal_huang} studied the optimal transmit power control strategies for AirComp. 
As an extension, the authors in \cite{chen2018uniform} investigated AirComp in multiple-input single-output (MISO) wireless networks, where a semi-definite relaxation (SDR) based successive convex approximation (SCA) algorithm was proposed to design the receive beamforming vector at the access point (AP). 
The authors in \cite{multi_function} and \cite{Muti_modal} integrated multiple-input multiple-output (MIMO) with AirComp, and studied the transceiver design for multi-function computation and multi-modal sensing, respectively.
The approximated receive beamformer was designed by utilizing the Grassman manifold theory in \cite{Muti_modal}.
However, the existing studies based on SDR and SDR-based SCA can only obtain sub-optimal solutions. 
The optimal receive beamforming design for AirComp in MISO systems is still not available in the literatures.
%

In this paper, we consider wireless data aggregation via AirComp in IoT networks with a multi-antenna AP.
Our goal is to minimize the computation distortion at the AP by jointly optimizing the transmit scalars at the transmitter and denoising factor and receive beamforming vector at the AP.
With the transmit scalars and the denoising factor derived in closed-form, the distortion minimization problem turns to a non-convex quadratically constrained quadratic programming (QCQP) problem with respect to the receive beamforming vector at the AP.
We propose a globally optimal branch and bound (BnB) algorithm to design the receive beamforming vector, thereby further reducing the distortion of AirComp when compared to the baseline algorithms, as verified via extensive simulations.
Moreover, the proposed algorithm can be treated as a benchmark to evaluate the quality of the solutions returned by the existing algorithms, e.g., SDR and SDR-based SCA.


\emph{Notations}: 
We use boldface upper-case, boldface lower-case, and lower-case letters to denote matrices, vectors, and scalars, respectively.
We denote the imaginary unit of a complex number as $\mathbf{j}$.
  $(\bm \cdot)^{\sf{H}}$ stands for conjugate transpose of a matrix or a vector.
 $\|\cdot\|$ denotes the $l_2$ norm operator.
 $\operatorname{Re}\{\bm\cdot\}$, $\operatorname{Im}\{\bm \cdot\}$, $|\cdot|$, and $\operatorname{arg}(\cdot)$ represent the real part, imaginary part, absolute value, and argument of a scalar, respectively.
 $\mathbb{E}\left[\bm \cdot\right]$ denotes the expectation of a random variable.
%

\section{System Model and Problem Formulation}

\subsection{System Model}
We consider fast wireless data aggregation via AirComp in an IoT system consisting of $K$ single-antenna IoT devices and one AP with $N$ antennas. We denote $\mathcal{K}=\{1,2,\ldots, K\}$ as the index set of IoT devices.
 The AP aims to recover the arithmetic mean of the sensory data from all IoT devices.
We denote $s_k = \varphi_k(z_k)$ as the transmit signal of device $k$, where $\varphi_k(\cdot)$ is the specific pre-processing function and
  $z_k\in\mathbb{C}$ is the representative information-bearing data at device $k$.
Without loss of generality,
we assume that $\{s_k\}_{k=0}^{K}$ are independent and have zero mean and unit power, i.e., $\mathbb{E}[s_{k}s_{k}^{\sf H}] = 1$, and $\mathbb{E}[s_{k}s_{j}^{\sf H}] = 0, \forall  k \neq j$ \cite{optimal_huang}.
To recover the arithmetic mean of the sensory data from all IoT devices, i.e., $\frac{1}{K}\sum_{k\in\mathcal{K}}z_k$, it is sufficient for the AP to estimate the following target function
\begin{align}
	g=\sum_{k\in\mathcal{K}}s_k.
\end{align}
By calibrating the transmission timing of each IoT device, we assume that the signals transmitted by all IoT devices are synchronized when receiving at the AP.
The signal received at the AP can be expressed as
\begin{align}
\bm{y}=\sum_{k\in\mathcal{K}}\bm{h}_{k}{w}_ks_k+\bm{n},
\end{align}
where $w_k\in\mathbb{C}$ denotes the transmit scalar of device $k$, $ \bm h_{k}\in\mathbb{C}^{N\times1}$ is the channel coefficient vector of the link from device $k$ to the AP,
 and $ \bm n\sim\mathcal{CN}(0, \sigma^2\bm I_N) $ is the additive white Gaussian noise (AWGN) with zero mean and variance $\sigma^2$. 
In practice, the maximum transmit power is limited, i.e., $|w_k|^2 \leq P, \forall k$.
After applying the receive combining, the estimated function at the AP is given by
\begin{align}
\hat{g}&={1\over{\sqrt \eta}}{\bm{m}^{\sf{H}}\bm{y}} =\!{1\over{\sqrt \eta}}{\bm{m}}^{\sf{H}}\sum_{k\in\mathcal{K}}\bm{h}_{k} {w}_ks_k +{1\over{\sqrt \eta}}\bm{m}^{\sf{H}}\bm{n},
\end{align} 
where $\bm{m}\in\mathbb{C}^N$ and $\eta$ denote the receive beamforming vector and the denoising factor at the AP, respectively.

\subsection{Problem Formulation}
To evaluate the performance of AirComp,
we adopt mean-squared-error (MSE) to quantify the distortion of $\hat{g}$ with respect to $g$, given by
\begin{align}
{\sf{MSE}}(\hat{g}, g)=\mathbb{E}\left(|\hat{g}-g|^2\right) = \!\! \sum_{k\in\mathcal{K}}\left|   \frac{{{\bm{m}}^{\sf{H}}\bm{h}_{k}{w}_k}}{\sqrt{\eta}} -1\right|^2 \!+\! \frac{\sigma^2\|\bm{m}\|^2}{\eta}\nonumber . 
\end{align}
When the receive beamforming  vector $\bm{m}$ is given, the optimal transmit scalars that minimize the MSE can be expressed as \cite{yangkai,chen2018uniform}
\begin{align}\label{a}
w_k^{\star}=\sqrt{\eta}{{(\bm{m}^{\sf{H}}\bm{h}_{k})^{\sf{H}}}\over{\|\bm{m}^{\sf{H}}\bm{h}_{k}\|^2}},\forall k. 
\end{align}
Due to the transmit power constraint, $\eta$ can be expressed as \begin{align}\label{b}
\eta=P\min_{k\in\mathcal{K}} \|\bm{m}^{\sf{H}}\bm{h}_{k}\|^2.
\end{align}
With \eqref{a} and \eqref{b}, the MSE  can be further rewritten  as
\begin{align}
{\sf{MSE}}={{\|\bm{m}\|^2\sigma^2}\over{\eta}}
={{\|\bm{m}\|^2\sigma^2}\over{P\min_{k\in\mathcal{K}} \|\bm{m}^{\sf{H}}\bm{h}_{k}\|^2}}.\nonumber
\end{align}
We thus propose to optimize the receive beamforming  vector $ \bm m $ to minimize the MSE as follows:  
\begin{equation}\label{eq:ori}
        \begin{split}
                \underset{\bm m}{\min} \left({{\|\bm{m}\|^2\sigma^2}\over{P\min_{k\in\mathcal{K}} \|\bm{m}^{\sf{H}}\bm{h}_{k}\|^2}}\right).
        \end{split}
\end{equation}
According to \cite{chen2018uniform}, problem \eqref{eq:ori} can be further equivalently transformed to the following problem
\begin{equation}\label{new} 
        \begin{split}
                \underset{\bm m}{\min}  &\quad \|\bm m\|^2\\
                \text{s.t.} &\quad\|\bm m^{\sf H}\bm{h}_{k}\|^2 \geq 1, ~  \forall k.
        \end{split}
\end{equation}
To this end, we formulate the MSE minimization problem as a non-convex QCQP problem.
The authors in \cite{chen2018uniform,jiang2019} solved the non-convex QCQP problem by proposing the SDR and SDR-based SCA algorithms, which, however, are sub-optimal.  
The quality of the solutions obtained by the aforementioned sub-optimal algorithms is still unknown due to the lack of the optimal algorithm. 
In the next section, we shall propose a globally optimal algorithm for the optimization of receive beamforming vector $\bm m$ to fully exploit the potential of multiple antennas and to evaluate the performance of the existing sub-optimal algorithms.

\section{Proposed Global Optimal BnB Algorithm}

The BnB algorithm
is capable of approaching an optimal solution within any desired error bound for some non-convex problems \cite{ChengLu}.
The main idea of the BnB algorithm is to first construct the lower bound and upper bound for the non-convex problem, and then lift the lower bound and reduce the upper bound iteratively
through judiciously designing a branching strategy.
Specifically, the lower bound can be obtained by solving a corresponding relaxation problem.
Subsequently, we project the solution of the aforementioned relaxation problem to the original feasible region to form an upper bound.


\subsection{Lower Bound and Upper Bound} 

\begin{figure}[t]
\centering
\includegraphics[width=2.5in]{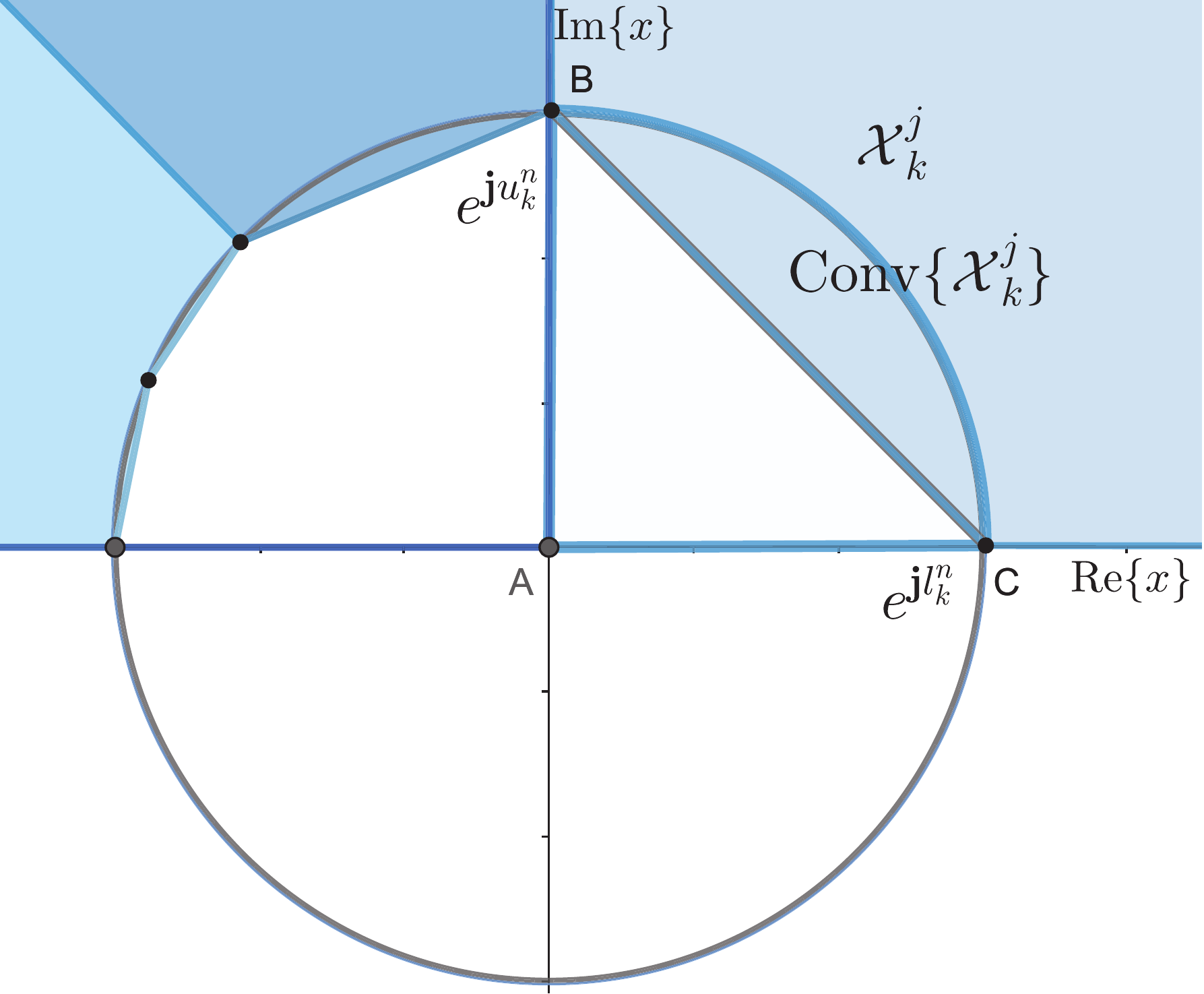}
\caption{Illustration of convex relaxation of the outer regions of arcs for three different argument intervals, i.e., $\pi/2$, $\pi/4$, and $\pi/8$.}
\label{demo}
\vspace{-5mm}
\end{figure}

To facilitate the BnB algorithm design, we first introduce an auxiliary variable $\bm{x} = [x_1,x_2,\ldots,x_K]^{\sf T} \in \mathbb{C}^{K}$, and then rewrite problem \eqref{new} as
\begin{equation}\label{ori:BnB}
        \begin{split}
                \underset{\bm m, \bm x}{\min}  &\quad \|\bm m\|^2\\
                \text{s.t.} &\quad \bm m^{\sf H}\bm{h}_{k} = x_k, ~ \forall k, \\
                &\quad |x_k|\geq1, ~ \forall k, 
        \end{split}
\end{equation}
where constraint $|x_k|\geq1$ means that the feasible region of $x_k$ is the outer region of the unit circle in a complex plane.
We denote set $\mathcal{X} = \{\bm x \big | |x_k|\geq1, ~ \forall k\}$,
which can be treated as the Cartesian product of $K$ sets, i.e., $\mathcal{X} = \prod_{k=1}^{K} \mathcal{X}_k$ where $\mathcal{X}_k = \left \{x_k\in \mathbb{C} \big | |x_k| \geq 1\right\},~ \forall k$.
For non-convex set $\mathcal{X}_k$, the corresponding convex hull is the whole complex plane.
However, such a relaxation is too loose to generate an effective lower bound.
To this end, we partition $\mathcal{X}_k$ into several subregions, leading to a tighter relaxation.
Specifically, for the $n$-th non-convex subregion 
$\mathcal{X}_k^n = \left\{x_k\in \mathbb{C} \Big | |x_k| \geq 1, ~ \arg (x_k) \in \left[l_k^n, u_k^n\right) \right\}$ with the argument interval being not greater than $\pi$, i.e., $u_k^n-l_k^n\leq \pi$, the corresponding convex hull can be represented as 
\begin{equation}
\label{convex_hull}
\begin{aligned}
&~~~\operatorname{Conv}\{\mathcal{X}_k^n\} \\
&= \left\{x_k\in \mathbb{C} \Big |
\begin{split}
 \operatorname{Re}\left\{\bar{x}_k \cdot \frac{e^{\mathbf{j} u_k^n}+e^{\mathbf{j} l_k^n}}{2}\right\} & \geq \cos \left(\frac{u_k^n-l_k^n}{2}\right), \\
\arg \left(x_k\right) & \in \left[l_k^n, u_k^n\right)
 \end{split}
 \right\}, 
\end{aligned}	
\end{equation}
where $\bar{x}_k$ denotes the conjugate of $x_k$.
For example, as shown in Fig. \ref{demo}, the convex hull is enclosed by the line BC between points $e^{\mathbf{j}l_k^n}$ and $e^{\mathbf{j}u_k^n}$ at the unit circle
\[
\left\{x_k\in \mathbb{C} \Big |
\begin{aligned}
\operatorname{Re}\left\{\bar{x}_k \cdot \frac{e^{\mathbf{j} u_k^n}+e^{\mathbf{j} l_k^n}}{2}\right\}  = \cos \left(\frac{u_k^n-l_k^n}{2}\right)
\end{aligned}
\right\},
\]
ray AB $\{x_k \in \mathbb{C} \mid \text{arg}(x_k) = l_k^n\}$, and ray AC $\{x_k \in \mathbb{C} \mid \text{arg}(x_k) = u_k^n\}$. 

\begin{remark}
\emph{The minimum modulus among the convex hull of $\mathcal{X}_k^n$, i.e., $\min_{x_k \in \operatorname{Conv}\{\mathcal{X}_k^n\}} |x_k|$, is $\cos \left(\frac{u_k^n-l_k^n}{2}\right)$ that corresponds to the middle point of line segment BC is also the furthest point to set $\mathcal{X}_k^n$.
The convex relaxation $\operatorname{Conv}\{\mathcal{X}_k^n\}$ approaches to $\mathcal{X}_k^n$ when $u_k^n-l_k^n$ approaches to zero. 
}	
\end{remark}

In the $t$-th iteration of the BnB algorithm, the original feasible region $\mathcal{X}$ is divided into several subregions $\{\mathcal{S}^i\}_{i\in \mathcal{I}_t}$, where $\mathcal{I}_t$ denotes the index set of subregions at the $t$-th iteration.
Specifically, we rewrite $\mathcal{S}^{i}$ as the Cartesian product of $K$ independent sets, i.e., $\mathcal{S}^{i} = \mathcal{X}^i_1\times \mathcal{X}^i_2\times \cdots \times \mathcal{X}^i_K$, 
where 
\[
\mathcal{X}^i_k = \left\{ x_k \Big |
\begin{aligned}
|x_k| = 1,
\arg \left(x_k\right) \in \left[ l^i_k, u^i_k \right)
\end{aligned}
\right\}, \forall k.
\]
Besides, we have $\cup_{i\in \mathcal{I}_t} \mathcal{S}^i = \mathcal{X}$ and $\mathcal{S}^i \cap\mathcal{S}^{i^{\prime}} = \emptyset, i \neq i', \forall i, i^{\prime} \in \mathcal{I}_t$.
As a result, problem \eqref{ori:BnB} can be separated into a series of subproblems $\{\mathcal{P}^i\}_{i\in \mathcal{I}_t}$ defined on subregions $\{\mathcal{S}^i\}_{i\in \mathcal{I}_t}$ as follows,
\begin{equation}\label{subproblem_i}
        \mathcal{P}^i : \quad \begin{aligned}
                \underset{\bm m, \bm x}{\min}  &\quad \|\bm m\|^2\\
                \text{s.t.} &\quad \bm m^{\sf H}\bm{h}_{k} = x_k, ~ \forall k, \\
                &\quad \bm{x} \in \mathcal{S}^i.
        \end{aligned}
\end{equation}

To obtain a lower bound for problem \eqref{subproblem_i},
we resort to solve its convex relaxation problem as follows,
\begin{equation}\label{subproblem:convex}
        \begin{split}
                \underset{\bm m, \bm x}{\min}  &\quad \|\bm m\|^2\\
                \text{s.t.} &\quad \bm m^{\sf H}\bm{h}_{k} = x_k, ~ \forall k, \\
                &\quad \bm{x} \in \hat{\mathcal{S}}_i,
        \end{split}
\end{equation}
where $\hat{\mathcal{S}}_i$ denotes the convex hull of $\mathcal{S}^i$.
Specifically, 
$\hat{\mathcal{S}}_{i} = \hat{\mathcal{X}}^i_1\times \hat{\mathcal{X}}^i_2\times \cdots \times \hat{\mathcal{X}}^i_K$, where $\hat{\mathcal{X}}^i_k$ is convex hull of $\mathcal{X}^i_k,~\forall k$.
It is worth noting that the convex hull of $\mathcal{X}^i_k$ can be obtained by using \eqref{convex_hull} if its argument interval is less than or equal to $\pi$, i.e., $u^i_k-l^i_k \leq \pi$.
The optimal objective value of convex problem \eqref{subproblem:convex} serves as a lower bound  for problem \eqref{subproblem_i} since $\mathcal{S}^i \subseteq \hat{\mathcal{S}}_i$.
We then take the minimum lower bound among $\{\mathcal{P}^i\}_{i\in \mathcal{I}_t}$ as the current lower bound of problem \eqref{ori:BnB}, denoted as $L^t$.

\begin{remark}
\emph{
The optimal solution of problem \eqref{ori:BnB} lies in one of $\{\mathcal{S}^i\}_{i\in\mathcal{I}_t}$ since $\cup_{i\in \mathcal{I}_t} \mathcal{S}^i = \mathcal{X}$. 
We denote the index of the subregion that incorporates the optimal solution as $i^{\prime}$.
The optimal objective value of $\mathcal{P}_{i^{\prime}}$ is identical to the optimal objective value of problem \eqref{ori:BnB}.
Therefore, the lower bound of $\mathcal{P}_{i^{\prime}}$ is less than the optimal objective value of problem \eqref{ori:BnB}.
However, it is challenging to identify which subregion the optimal solution lies in.  
Fortunately, the minimum lower bound among $\{\mathcal{P}^i\}_{i\in \mathcal{I}_t}$ will not be larger than the lower bound of $\mathcal{P}_{i^{\prime}}$. 
As a result, the minimum lower bound among $\{\mathcal{P}^i\}_{i\in \mathcal{I}_t}$ can serve as a lower bound of problem \eqref{ori:BnB}.
}	
\end{remark}

On the other hand, the objective value of problem \eqref{subproblem_i} at any point located in feasible region $\mathcal{S}^i$ can serve as its upper bound.
We scale the optimal solutions of problem \eqref{subproblem:convex}, denoted as $\bm{x}^i_*$ and $\bm{m}^i_*$ to generate a point that belongs to $\mathcal{S}^i$ as follows
\begin{equation}
\label{upper_bound}
\begin{aligned}
	\tilde{\bm{x}}^i_{*} &= \frac{\bm{x}^i_*}{\min\{|(x^i_*)_1|,|(x^i_*)_2|,\ldots,|(x^i_*)_K|,1\}} \in \mathcal{S}^i, \\
	\tilde{\bm{m}}^i_{*} &= \frac{\bm{m}^i_*}{\min\{|(x^i_*)_1|,|(x^i_*)_2|,\ldots,|(x^i_*)_K|,1\}},
\end{aligned}	
\end{equation}
where $(x^i_*)_k,~\forall k$ denotes the $k$-th element of $\bm x^i_*$.
As a result, $\|\tilde{\bm{m}}^i_*\|^2$ can be treated as an upper bound of problem \eqref{subproblem_i}.
The upper bound of problem \eqref{ori:BnB}, denoted as $U^t$, can be updated by the minimum upper bound among the current problem set $\{\mathcal{P}^i\}_{i\in \mathcal{I}_t}$.

\begin{remark}
\emph{All the points in $\{\tilde{\bm{x}}^i_*\}_{i\in \mathcal{I}_t}$ belong to the feasible region of problem \eqref{ori:BnB}.
As a result, all of the corresponding objective values can serve as an upper bound of problem \eqref{ori:BnB}.
To construct a tighter upper bound for problem \eqref{ori:BnB}, we take the minimum upper bound of $\{\mathcal{P}^i\}_{i\in \mathcal{I}_t}$ to be the upper bound of problem \eqref{ori:BnB}.
}	
\end{remark}

\subsection{Branching Strategy}

By performing partition on the feasible regions of current subproblems, we can get more subproblems with smaller feasible regions.
The corresponding relaxation become tighter as the partition continues, and the gap between the upper bound and lower bound diminishes.
On the other hand, $\min\{|(x^i_*)_1|,|(x^i_*)_2|,\ldots,|(x^i_*)_K|\}$ will increase as the relaxations become tighter.
As a result,  
according to \eqref{upper_bound}, 
the upper bound of problem \eqref{ori:BnB} will decrease as the partition continues.

 
Specifically, in the $t$-th BnB iteration, we shall select a problem with the minimum lower bound in the problem set $\{\mathcal{P}^i\}_{i\in \mathcal{I}_t}$ and perform subdivision on its feasible region. 
Without loss of generality, we denote the problem as $\mathcal{P}^{i_t}$, and the solution of the corresponding convex relaxation problem as $\bm{x}_{i_t}^{*}$.
For convenience of elaborating the partition rule, we rewrite $\mathcal{S}^{i_t}$ in the form of Cartesian product of $K$ independent parts, i.e., $\mathcal{S}^{i_t} = \mathcal{X}^{i_t}_1\times \mathcal{X}^{i_t}_2\times \cdots \times \mathcal{X}^{i_t}_K$.
Subsequently, we partition current region $ \mathcal{S}^{i_t}$ into two subregions, i.e., 
$\mathcal{S}^{i_t}_l = \mathcal{X}^{i_t}_1\times \mathcal{X}^{i_t}_2\times \cdots \times \left(\mathcal{X}^{i_t}_{k^{t}}\right)^l \times \cdots \times \mathcal{X}^{i_t}_K$
 and 
$\mathcal{S}^{i_t}_r = \mathcal{X}^{i_t}_1\times \mathcal{X}^{i_t}_2\times \cdots \times \left(\mathcal{X}^{i_t}_{k^{t}}\right)^r \times \cdots \times \mathcal{X}^{i_t}_K$, 
where $k^{t} = \argmin_{i}\{|(x^{i_t}_{*})_i|\}$.
The only difference between $\mathcal{S}^{i_t}_{l}$ and $\mathcal{S}^{i_t}_{r}$ is the $k^{t}$-th part, where the original region is divided into two equal parts, i.e., $\left(\mathcal{X}^{i_t}_{k^{t}}\right)^l$ and $\left(\mathcal{X}^{i_t}_{k^{t}}\right)^r$.
For instance, if $\mathcal{X}^{i_t}_{k^{t}} = 
\left\{x_{k^{t}} \Big |
\begin{aligned}
|x_{k^{t}}| = 1,
\arg \left(x_{k^{t}}\right)  \in \left[ l^{i_t}_{k^{t}},u^{i_t}_{k^{t}}\right)
\end{aligned}
\right\}$, then
\begin{subequations} \label{partition}
\begin{align}
\left(\mathcal{X}^{i_t}_{k^{t}}\right)^l &= 
\left\{x \Big |
\begin{aligned}
|x_{k^{t}}| = 1,
\arg \left(x\right)  \in \left[ l^{i_t}_{k^{t}}, \frac{l^{i_t}_{k^{t}}+u^{i_t}_{k^{t}}}{2} \right)
\end{aligned}
\right\}, \\
\left(\mathcal{X}^{i_t}_{k^{t}}\right)^r &= 
\left\{x_{k^{t}} \Big |
\begin{aligned}
|x_{k^{t}}| = 1,
\arg \left(x_{k^{t}}\right)  \in \left[\frac{l^{i_t}_{k^{t}}+u^{i_t}_{k^{t}}}{2}, u^{i_t}_{k^{t}} \right)
\end{aligned}
\right\}.
\end{align}	
\end{subequations}

As a result, problem $\mathcal{P}^{i_t}$ is branched into the following two subproblems
\begin{equation}
\begin{aligned}
\mathcal{P}^{i_t}_l: ~	\underset{\mathbf{m}, \bm{x}}{\min}  &\quad \|\bm m\|^2\\
    \text{s.t.} &\quad \bm m^{\sf H}\bm{h}_{k} = x_k, ~ \forall k, \\
    &\quad \bm x \in \mathcal{S}^{i_t}_{l}.
\end{aligned}
\end{equation}
\begin{equation}
\begin{aligned}
\mathcal{P}^{i_t}_r: ~	\underset{\mathbf{m}, \bm{x}}{\min}  &\quad \|\bm m\|^2\\
    \text{s.t.} &\quad \bm m^{\sf H}\bm{h}_{k} = x_k, ~ \forall k, \\
    &\quad \bm x \in \mathcal{S}^{i_t}_{r}.
\end{aligned}
\end{equation}

The lower bound and upper bound of problem $\mathcal{P}^{i_t}_l$ and $\mathcal{P}^{i_t}_r$ can be obtained according to the rules discussed in last subsection.
Finally, we add the two problems into the problem set $\{\mathcal{P}^i\}_{i\in\mathcal{I}_{t+1}}$ and remove $\mathcal{P}^{i_t}$ from it, where $\mathcal{I}_{t+1}$ is the updated index set of subregions at the $(t+1)$-th iteration. 

\subsection{Complexity}
With the aforementioned rules for constructing bounds and the branching strategy, the BnB algorithm is guaranteed to converge to an $\epsilon$-optimal solution within at most $\left(2 \pi/ \arccos \left(\frac{1}{\sqrt{1+\epsilon}}\right)\right)^{K}+1$ iterations \cite{ChengLu}.
Besides, in each iteration, the computation of the lower bound dominates the complexity of the proposed algorithm, which involves solving a convex QCQP problem, i.e., \eqref{subproblem:convex}.
According to \cite{Nesterov_interior}, the optimal solution for problem \eqref{subproblem:convex} can be obtained by using the standard interior-point method with complexity $\mathcal{O}(N^3K^{3.5})$.
As a result, the computation time complexity of the proposed BnB algorithm is 
$\mathcal{O}(TN^3K^{3.5})$, where $T = \left(2 \pi/ \arccos \left(\frac{1}{\sqrt{1+\epsilon}}\right)\right)^{K}+1$.

\begin{algorithm}[t]
	\caption{BnB Algorithm for Solving Problem \eqref{ori:BnB}}
	\begin{algorithmic}[1]
		\STATE Initialize $\mathcal{S}^0 = \prod_{i=1}^{K}[0,2\pi]$. 
		Randomly generate $\bm{m}$.
		Set $\bm{m}_{*} = \frac{\bm{m}}{\operatorname{max}_k\{|\bm m^{\sf H}\bm{h}_{k}|\}}$, $\bar{\bm{m}}_{*} = \frac{\bm{m}}{\operatorname{min}_k\{|\bm m^{\sf H}\bm{h}_{k}|\}}$.
		Set $(x^0_*)_k = \bm{m}_{*}\bm{h}_{k},~\forall k$.
		Lower bound $L_0$ and upper bound $U_0$ are set to be $\|\bm{m}_{*}\|^2$ and $\|\bar{\bm{m}}_{*}\|^2$, respectively.
		Use problem \eqref{ori:BnB} with $\{L_0,U_0, \bm x^0_*,\mathcal{S}^0\}$ to initialize problem set.
		Set convergence tolerance $\epsilon$ and iteration index $t=0$, 
		\REPEAT 
		\STATE Select problem $\mathcal{P}^{i_t}$ with the smallest lower bound among current problem set $\{\mathcal{P}^i\}_{i\in \mathcal{I}_t}$;
		\STATE Partition the feasible set of the selected problem into two subregions, $\mathcal{S}^{i_t}_\mathrm{l}$ and $\mathcal{S}^{i_t}_\mathrm{r}$, according to \eqref{partition};
		\STATE Compute the lower bound and upper bound for $\mathcal{P}^{i_t}_{l}$ and record the solutions;
		\STATE Compute the lower bound and upper bound for $\mathcal{P}^{i_t}_{r}$ and record the solutions;
		\STATE Add problems $\mathcal{P}^{i_t}_{l}$ and $\mathcal{P}^{i_t}_{r}$ to problem set $\{\mathcal{P}^i\}_{i\in\mathcal{I}_{t+1}}$;
		\STATE $t\leftarrow t+1$;
		\STATE Update upper bound $U^t$ and lower bound $L^t$ for problem \eqref{ori:BnB} as the smallest upper bound and lower bound among $\{\mathcal{P}^i\}_{i\in\mathcal{I}_{t}}$, respectively;
		\UNTIL $\frac{U^t-L^t}{L^t}\le\epsilon$
	\end{algorithmic}
\end{algorithm}

\section{Simulation Results}

 In this section, we present the simulation results of the proposed algorithm for AirComp in IoT networks. 
 We consider a three-dimentional setting, where the AP is located at $(0,0,20)$, while the IoT devices are uniformly located within a circular region centered at $(120,~ 20,~ 0)$ meters with radius $20$ meters. 
 The antennas at the AP are arranged as a uniform linear array. 
In the simulations, we consider both large-scale fading and small-scale fading for the channel.
The distance-dependent large-scale fading is modeled as $T_0(d/d_0)^{-\alpha}$, 
where $T_0$ is the path loss at the reference distance $d_0 = 1$ meter, $d$ denotes the distance between transmitter and receiver, and $\alpha$ is the path loss exponent.
Besides, we model the small-scale fading as Rician fading with rician factor $\beta$.
All results in the simulations are obtained by averaging over $500$ channel realizations. 
Unless specified otherwise, we set $\alpha = 3$, $T_0 = -30 $ dB, $\beta = 3$, $P = 30$ dBm, $\sigma^2 = -100$ dBm, and $\epsilon = 10^{-5}$.

\subsection{Convergence Performance}
\begin{figure*}[h]
        \centering
        \subfigure[MSE versus the number of iterations when $K=8$ and $N=4$.]{
        \label{convergence}
        \includegraphics[width=0.66\columnwidth]{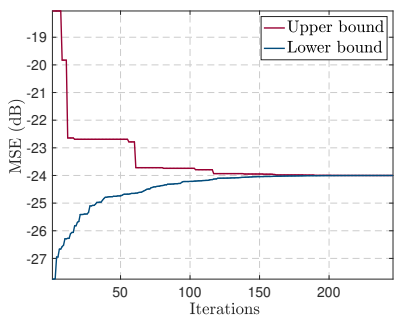}}
        \subfigure[MSE versus the number of antennas at AP when $K=10$.]{
        \label{antenna}
        \includegraphics[width=0.66\columnwidth]{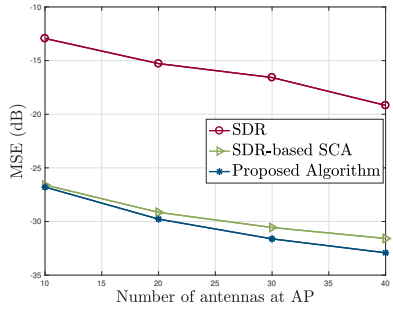}}
        \subfigure[MSE versus the number of IoT devices when $N=10$.]{
        \label{devices}
        \includegraphics[width=0.66\columnwidth]{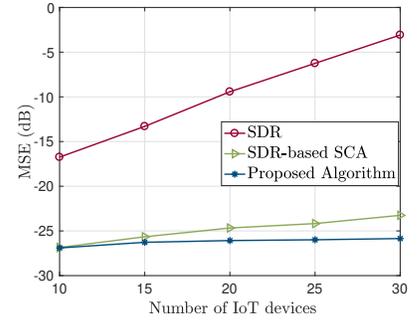}} 
        \caption{Performance of the proposed BnB algorithm for AirComp in IoT networks.}
\end{figure*}

We present the convergence performance of the proposed BnB algorithm in Fig. \ref{convergence}.
It can be observed that the upper bound decreases and the lower bound increases as the iteration proceeds.
In addition, the gap between the upper bound and the lower bound diminishes as the number of iterations increases.
In particular, the algorithm terminates within 250 iterations, where the gap between the upper bound and the lower bound is below a predefined convergence tolerance.

\subsection{Performance Evaluation of the Existing Algorithms}


In this subsection, we compare the proposed BnB algorithm with SDR \cite{Tom_luo_sdr} and SDR-based SCA \cite{chen2018uniform} algorithms.

Fig. \ref{antenna} shows the impact of the number of antennas at the AP on the MSE when the number of IoT devices $K = 10$.
As can be observed, the MSE of AirComp monotonically decreases as the number of antennas increases. 
This is because deploying a larger antenna array leads to a greater diversity gain.
Besides, it is clear that the proposed BnB algorithm has the best performance in minimizing the MSE.
This is because our proposed BnB algorithm is the global optimization algorithm that has the ability to approach the optimal solution within any desired error tolerance.
The performance gap between our proposed BnB algorithm and the SDR method is considerably large, as the SDR method is weak at optimizing the AirComp system.
By comparing the SDR-based SCA algorithm with the proposed algorithm, one can claim that the former can obtain a high-quality solution in the sense of the MSE.

The MSE versus the number of the IoT devices is plotted in Fig. \ref{devices}, where the number of antennas at the AP is set to be $10$.
It is obvious that the quality of the solutions of the SDR and SDR-based SCA algorithms degenerates as the number of IoT devices increases.
This is because the performance of the SDR-based SCA algorithm heavily depends on the quality of the solution returned by the SDR algorithm, which usually does not work well as the number of IoT devices increases.

\section{Conclusions}

In this paper, we investigated the joint design of the transmit scalars, the denoising factor, and the receive beamforming vector for AirComp in IoT networks.  
We derived the closed-form expressions for the transmit scalars and the denoising factor, resulting in a non-convex QCQP problem with respect to the receive beamforming vector at the AP.
We then proposed a global optimal BnB algorithm to optimize the receive beamforming vector.
The achieved MSE by the proposed BnB algorithm in the simulations revealed the substantial potential of optimizing the receive beamformer.
Our proposed algorithm can be adopted as a benchmark to evaluate the performance of the existing sub-optimal algorithms, e.g., SDR and SDR-based SCA.

%

\ifCLASSOPTIONcaptionsoff
  \newpage
\fi

\bibliographystyle{IEEEtran}  
\bibliography{reference.bib}

\begin{thebibliography}{10}
\providecommand{\url}[1]{#1}
\csname url@samestyle\endcsname
\providecommand{\newblock}{\relax}
\providecommand{\bibinfo}[2]{#2}
\providecommand{\BIBentrySTDinterwordspacing}{\spaceskip=0pt\relax}
\providecommand{\BIBentryALTinterwordstretchfactor}{4}
\providecommand{\BIBentryALTinterwordspacing}{\spaceskip=\fontdimen2\font plus
\BIBentryALTinterwordstretchfactor\fontdimen3\font minus
  \fontdimen4\font\relax}
\providecommand{\BIBforeignlanguage}[2]{{%
\expandafter\ifx\csname l@#1\endcsname\relax
\typeout{** WARNING: IEEEtran.bst: No hyphenation pattern has been}%
\typeout{** loaded for the language `#1'. Using the pattern for}%
\typeout{** the default language instead.}%
\else
\language=\csname l@#1\endcsname
\fi
#2}}
\providecommand{\BIBdecl}{\relax}
\BIBdecl

\bibitem{zhu2020overtheair}
\BIBentryALTinterwordspacing
G.~Zhu, J.~Xu, K.~Huang, and S.~Cui, ``Over-the-air computing for wireless data
  aggregation in massive {IoT},'' 2020. [Online]. Available:
  \url{https://arxiv.org/abs/2009.02181}
\BIBentrySTDinterwordspacing

\bibitem{blind}
J.~{Dong}, Y.~{Shi}, and Z.~{Ding}, ``Blind over-the-air computation and data
  fusion via provable wirtinger flow,'' \emph{IEEE Trans. Signal Process.},
  Jan. 2020.

\bibitem{wang_iot}
Z.~Wang, Y.~Shi, Y.~Zhou, H.~Zhou, and N.~Zhang, ``Wireless-powered
  over-the-air computation in intelligent reflecting surface-aided {IoT}
  networks,'' \emph{IEEE Internet Things J.}, vol.~8, no.~3, pp. 1585--1598,
  Feb. 2021.

\bibitem{wang2020federated}
\BIBentryALTinterwordspacing
Z.~Wang, J.~Qiu, Y.~Zhou, Y.~Shi, L.~Fu, W.~Chen, and K.~B. Lataief,
  ``Federated learning via intelligent reflecting surface,'' 2020. [Online].
  Available: \url{https://arxiv.org/abs/2011.05051}
\BIBentrySTDinterwordspacing

\bibitem{yangkai}
K.~{Yang}, T.~{Jiang}, Y.~{Shi}, and Z.~{Ding}, ``Federated learning via
  over-the-air computation,'' \emph{IEEE Trans. Wireless Commun.}, vol.~19,
  no.~3, pp. 2022--2035, Mar. 2020.

\bibitem{info}
B.~{Nazer} and M.~{Gastpar}, ``Computation over multiple-access channels,''
  \emph{IEEE Trans. Inf. Theory}, vol.~53, no.~10, pp. 3498--3516, Oct. 2007.

\bibitem{optimal_liu}
W.~{Liu}, X.~{Zang}, Y.~{Li}, and B.~{Vucetic}, ``Over-the-air computation
  systems: Optimization, analysis and scaling laws,'' \emph{IEEE Trans.
  Wireless Commun.}, vol.~19, no.~8, pp. 5488--5502, Aug. 2020.

\bibitem{optimal_huang}
X.~{Cao}, G.~{Zhu}, J.~{Xu}, and K.~{Huang}, ``Optimized power control for
  over-the-air computation in fading channels,'' \emph{IEEE Trans. Wireless
  Commun.}, vol.~19, no.~11, pp. 7498--7513, Nov. 2020.

\bibitem{chen2018uniform}
L.~{Chen}, X.~{Qin}, and G.~{Wei}, ``A uniform-forcing transceiver design for
  over-the-air function computation,'' \emph{IEEE Wireless Commun. Lett.},
  vol.~7, no.~6, pp. 942--945, Dec. 2018.

\bibitem{multi_function}
L.~{Chen}, N.~{Zhao}, Y.~{Chen}, F.~R. {Yu}, and G.~{Wei}, ``Over-the-air
  computation for {IoT} networks: Computing multiple functions with antenna
  arrays,'' \emph{IEEE Internet Things J.}, vol.~5, no.~6, pp. 5296--5306, Jun.
  2018.

\bibitem{Muti_modal}
G.~{Zhu} and K.~{Huang}, ``{MIMO} over-the-air computation for high-mobility
  multimodal sensing,'' \emph{IEEE Internet Things J.}, vol.~6, no.~4, pp.
  6089--6103, Aug. 2019.

\bibitem{jiang2019}
T.~{Jiang} and Y.~{Shi}, ``Over-the-air computation via intelligent reflecting
  surfaces,'' in \emph{Proc. IEEE Global Commun. Conf. (Globecom)}, Waikoloa,
  HI, Dec. 2019.

\bibitem{ChengLu}
C.~{Lu} and Y.~{Liu}, ``An efficient global algorithm for single-group
  multicast beamforming,'' \emph{IEEE Trans. Signal Process.}, vol.~65, no.~14,
  pp. 3761--3774, Jul. 2017.

\bibitem{Nesterov_interior}
Y.~Nesterov and A.~Nemirovskii, \emph{Interior-Point Polynomial Algorithms in
  Convex Programming}.\hskip 1em plus 0.5em minus 0.4em\relax Soc. Ind. Appl.
  Math., 1994.

\bibitem{Tom_luo_sdr}
Z.~{Luo}, W.~{Ma}, A.~M. {So}, Y.~{Ye}, and S.~{Zhang}, ``Semidefinite
  relaxation of quadratic optimization problems,'' \emph{IEEE Signal Process.
  Mag.}, vol.~27, no.~3, pp. 20--34, May 2010.

\end{thebibliography}

\end{document}